\documentclass[english,keywords,amsmath,amssymb,twocolumn]{revtex4}
\usepackage[T1]{fontenc}
\usepackage[latin1]{inputenc}
\usepackage{braket}
\usepackage{babel}%
\usepackage{graphicx}
\usepackage{color}
\usepackage{bm}
\usepackage{longtable}
\usepackage{amsmath}
\usepackage{amsfonts}
\usepackage{dsfont}
\usepackage{amssymb}
\usepackage{hyperref}
\begin{document}
\title{Interference experiment, anomalous weak value and Leggett-Garg test of macrorealism}
\author{A. K. Pan \footnote{akp@nitp.ac.in}}
\affiliation{National Institute of Technology Patna, Ashok Rajhpath, Patna 800005, India}
\begin{abstract}
Macrorealism is a classical world view asserting that the properties of macro-objects exist independently and irrespective of observation. One practical approach to test this view in quantum theory is to observe the quantum coherence for macro-object in an interference experiment. An elegant and conceptually appealing  approach for testing the notion of macrorealism in quantum theory is  through the violation of Leggett-Garg inequality. However, a conclusive Leggett-Garg test hinges on  how the non-invasive measurability criteria is guaranteed in an experiment and remains a debated issue till date. In this work, we connect the practical and the conceptual approaches for testing the macrorealism through the weak value. We argue that whenever quantum effect is observed in an interference experiment there is an existence of anomalous weak value. Further, we demonstrate that whenever such weak value exists, one obtains the violation of a Leggett-Garg inequality in any interference experiment. Since the system itself serves as apparatus in a path-only interference experiment, one cannot salvage the macrorealism by abandoning the non-invasive measurability assumption. We provide a rigorous discussion about the assumptions involved in Leggett-Garg scenario and how our scheme fits into it.  Further, using the quasiprobability approach we explain how non-invasive measurability condition  is satisfied in our scheme. 
\end{abstract}
\maketitle
\section{Introduction}
Macrorealism is a belief that the properties of a macroscopic object in our everyday-world must always possess definite ontic state, even though the state is not precisely known. Since the inception of quantum mechanics (QM), it remains a debatable question as to how such a view can be reconciled with the formalism of QM. Historically, this issue was first raised by Schr$\ddot{o}$dinger \cite{sch} through his famous thought experiment. Since then, quite a number of attempts have been made to pose the appropriate questions relevant to this issue and to answer those questions. An approach within the formalism of QM is the decoherence program \cite{zur} which explains how interaction between quantum systems and environment leads to classical behavior, but does not by itself provide the desired `cut' (\emph{\`a la} Heisenberg \cite{He25}). Even when the decoherence effect is made negligible, the quantum effect may disappear by coarsening the measurements\cite{bruk}.  An unified description of microscopic and macroscopic systems was also   provided \cite{ghi} by suitably modifying  the dynamics of standard QM. 

There is yet another approach to examine the macro-objectification problem through the realization of quantum coherence of the Schr$\ddot{o}$dinger cat-like states for large objects in an interference experiments \cite{arndt,gerlich}. This is in fact a practical approach to test the validity of QM for large object. Arndt \emph{et al.,} \cite{arndt} first experimentally produced the  double-slit interference of $C_{60}$ molecule having mass $720 \ amu$. The largest molecule till date that exhibits interference pattern is TPPF152 having mass $5310 \ amu$ and size $500 \ nm$ \cite{gerlich}. An important question in this context is how such experiments shed lights on the subtle conceptual notion macrorealism. 

However, the aforementioned attempts do not directly address the fundamental question whether macrorealism is, in principle, compatible with the formalism of QM. In 1985, Leggett and Garg  \cite{leggett85} first provided a refined definition of macrorealism. Specifically, the notion of macrorealism advocated by them consists of two main assumptions \cite{leggett85,leggett02} are the following.  (i) \emph{Macrorealism }{\emph per se} (MRps): If a macroscopic system has two or more states available to it, it remains in one of them at any instant of time. (ii) \emph{Non-invasive measurability (NIM)}: It is possible, at least in principle, to determine the  state of the system without affecting the system itself and its subsequent dynamics. Now, consider that  the measurement of a dichotomic  observable $\hat{M}$ is performed at three different times $t_1$, $t_2$ and $t_3$ $ (t_3 \geq t_2 \geq t_1 )$. In Heisenberg picture, this in turn implies the sequential measurement of the observables  $\hat{M}_{1},\hat{M}_{2}$ and $\hat{M}_{3}$ corresponding to $t_1$, $t_2$ and $t_3$ respectively. Using the above two assumptions, one can derive the standard Leggett-Garg inequality (LGI) 
\begin{eqnarray}
\label{eq1}
K_{3}&=&m_1 m_2\langle \hat{M_{1}} \hat{M_{2}}\rangle + m_2 m_3\langle \hat{M_{2}} \hat{M_{3}}\rangle\\
\nonumber
& - &m_1 m_3\langle \hat{M_{1}} \hat{M_{3}}\rangle -1 \leq {0} 
\end{eqnarray}
where $m_i m_j=\pm 1$ and $\langle {M_{i}} {M_{j}}\rangle=\sum\limits_{m_{i},m_{j}=\pm1} m_{i} m_{j} p(m_{i}, m_{j})$ with $i,j=1,2,3$ and $i<j$. This inequality is violated for certain state and measurement even for a qubit system. Such a violation is supposed to establish that a system is not behaving macrorealistically. 

We note here that there are two separate programs for the similar task. The purpose of practical approach is to test macroscopic quantum coherence and the goal of the conceptual approach to demonstrate the incompatibility between macrorealism and QM through the violation LGIs. While the former wishes to establish the quantumness of macro objects, the later demonstrates the prohibition of classicality of macro objects. Hence, an important question would be to ask here whether one implies other in some way. We demonstrate here that the interference experiments for testing macroscopic quantum coherence itself constitute a test of LGIs.       
 
There is a common perception that LGI is \emph{temporal analog} of Bell's  inequalities. This inference is motivated from the structural resemblance between Bell-CHSH inequalities and four-time LGIs. However, LG test shows serious pathology \cite{wild,clemente15,clemente16,halli16,swati17,halli17,pan17,kumari18,pan18,halli19,halli19a} in comparison to a Bell test. The non-signaling in space condition (the statistical version of locality postulate) is always satisfied in any physical theory. Thus, the violation of a given Bell's local realist inequality can unequivocally warrant a failure of local realism and not the locality alone. In contrast, the statistical version of non-invasive measurability assumption, the operational non-invasiveness (well-known as no-signalling in time) is in general not satisfied in QM. Thus, rather than attributing the violation of LGIs as an incompatibility of macrorealism with QM,  a stubborn macrorealist can always salvage the macrorealism in QM by simply abandoning non-invasive measurability. Hence to draw meaningful conclusions from a LG test one requires the satisfaction of no-signaling in time condition in QM by adopting a suitable measurement scheme. 

In their original paper,  Leggett and Garg proposed ideal negative-result measurements  to ensure the non-invasive measurability in QM which is adopted in recent experiments \cite{knee,robens}. However, this approach is criticized by pointing out that collapse indeed occurs in negative-result measurement procedure and eventually disturbs the system and its subsequent dynamics \cite{maroney}. Another approach to tackle the non-invasive measurability in experiments is by using the technique of weak measurement \cite{aav}. In weak measurement the strength of the measurement is possible to adjust and, in principle, the back action (the invasiveness) on the system can be reduced to an arbitrarily small amount \cite{pala,goggin,avella}. An interesting approach to tackle this issue was provided in \cite{halli16} by using suitable quasiprobabilities which can be positive or negative. By construction, they provide correct sequential correlation and importantly satisfy non-invasive measurability in QM. The set of quasiprobabilities, when positive, provides four two-time LGIs. Interestingly, when they are negative provide the violation of LGIs by keeping non-invasiveness satisfied. Another approach through the violation of an augmented set of LGIs using a noninvasive continuous-in-time velocity measurement has also been reported \cite{hali16aa,maji} 
 
It is an in$\acute{e}$luctable feature of measurement described by QM that it entails an interaction between the measuring apparatus and the observed system resulting in the state of the measured system to be necessarily entangled with the state of the observing apparatus. If the measurement is perfect then the system is disturbed the most. However, the non-classicality of a system (including macrosystem) can also be revealed even when system itself serves as a apparatus, for example, in the aforementioned interference experiments \cite{arndt,gerlich}. Suppose one wishes to measure the length of an one-dimensional object (system) in our everyday world. She then requires a scale (apparatus) whose pointer reading provides the length of that object.  However, measuring the length of the scale (becomes system now) the observer does not need to bring another scale (apparatus). Rather, the system itself serves as the role of apparatus and hence there is no way the entanglement occurs in QM and eventually system remain undisturbed by the measurement. The most famous example where a system behaves as apparatus is the double-slit experiment (in fact, any path-only interference experiment) where the interference pattern produces by matter waves are truly the quantum in nature. This provides a way to implement the non-invasive measurability condition in LG test. 

The purpose of this work is to provide a connection between the two approaches for testing macrorealism; the practical approach of interference experiment of large-object and the conceptual approach of LG test satisfying non-invasive measurebility. For this we demonstrate the following two theses. First, if a path-only interference experiment produces quantum effect (destructive interference) then there is an existence of anomalous weak value. Second, such existence of anomalous weak value warrants the incompatibility between macrorealism and QM through the violation of LGIs. Since the system itself serves  the role of apparatus and is not disturbed by the act of the measurement in a path-only interference experiment, the NIM condition is in principle satisfied.  We  also provide a discussion about the assumptions in LG scenario and how our scheme fits into it. Further, through quasiprobability approach \cite{halli16} we explain how the operational non-invasiveness (commonly known as no-signaling in time condition) is satisfied here. Thus, in our scheme macrorealism cannot be salvaged by abandoning NIM and hence it provides a decisive test of the notion of macrorealism LG alluded to.
 
 \section{Anomalous weak value and the violation of Leggett-Garg inequality} 

The anomalous weak value appears in a novel conditional measurement protocol first introduced in \cite{aav} widely known as weak measurement aided with post-selection (in short, weak measurement). Such conditional average value of an observable is experimentally measurable and can be beyond eigenvalue ranges of the concerned observable (for a review, see \cite{dre}). For simplicity, consider a two-dimensional system is prepared in a state $|i\rangle$ and the measured observable is denoted by $\hat{A}$. Let one wishes to collect the measurement statistics through another measurement basis corresponding to an observable $B=|f\rangle\langle f|-|f^{\prime}\rangle\langle f^{\prime}|$ where  $|f\rangle\langle f|+|f^{\prime}\rangle\langle f^{\prime}|=\mathbb{I}$ where $|f\rangle$ and $f^{\prime}\rangle$  are the eigenstates of that observable $B$. Then the expectation value can be written as 

\begin{align}
	\langle \hat{A}\rangle=\langle i| \hat{A}|i\rangle=\langle i|\hat{A}|f\rangle\langle f|i\rangle+\langle i|\hat{A}|f^{\prime}\rangle\langle f^{\prime}|i\rangle
\end{align}

Further re-arrangement provides us  

\begin{eqnarray}
\label{exp}
	\langle \hat{A}\rangle&=& |\langle f|i\rangle|^2 \frac{\langle i|\hat{A}|f\rangle}{\langle i|f\rangle}+ |\langle f^{\prime}|i\rangle|^2 \frac{\langle i|\hat{A}|f^{\prime}\rangle}{\langle i |f^{\prime}\rangle}\\
	\nonumber
	&=& p(f) (A)_{w}^{f} +p(f^{\prime}) (A)_{w}^{f^{\prime}}
\end{eqnarray}
 
where   $(A)_{w}^{f}= \langle i|\hat{A}|f\rangle/\langle i|f\rangle$ is the well-known definition of  weak value \cite{aav} and $p(f) $ is the post-selection probability. Similarly for the other term in Eq.(\ref{exp}). Thus, the expectation value can be calculated in a different measurement arrangement through the evaluation of weak values which are also experimentally measurable quantity. Note that, whenever  $\langle i|f\rangle$ is sufficiently small the weak value $(A)_{w}^{f}$ becomes anomalous, i.e., it can be beyond the ranges of eigenvalues. However, higher the weak value $(A)_{w}^{f}$ lower the post-selection probability $p(f) $. Note also that, both   $(A)_{w}^{f}$  and $(A)_{w}^{f^{\prime}}$ cannot be anomalous together. Another way to understand the connection between the expectation value and  weak value of $\hat{A}$ is that one wants to evaluate the expectation value through the suitable conditional sub-ensemble statistics. 

The three-time LGIs in Eq. (\ref{eq1}) can be cast into a two-time ones if the system is prepared in a particular initial state $|\psi_{i}\rangle=|+m_1\rangle$ where $\hat{M_{1}}|+m_1\rangle=|+m_1\rangle$, so that, $m_1=+1$. In that case we have

\begin{eqnarray}
\label{lgq}
(K_{3})_{Q}&=m_2 \langle\hat{M_{2}}\rangle_{+m_1\rangle} + m_2 m_3 \langle\hat{M_{2}}\hat{M_{3}}\rangle_{|+m_1\rangle} \\
\nonumber
& -  m_3\langle\hat{M_{3}}\rangle_{|+m_1\rangle} -1
\end{eqnarray}
In a macrorealistic model $K_{3}\leq 0$. For different values of $m_2=\pm1$ and $m_3=\pm1$ we have four two-time LGIs can be written as 
\begin{subequations}
\begin{eqnarray}
\label{tw1}
&&K_{31}=1-\langle\hat{M_{2}}\rangle -  \langle\hat{M_{2}}\hat{M_{3}}\rangle + \langle\hat{M_{3}}\rangle \geq 0\\
\label{tw2}
&&K_{32}=1+\langle\hat{M_{2}}\rangle +  \langle\hat{M_{2}}\hat{M_{3}}\rangle + \langle\hat{M_{3}}\rangle  \geq 0\\
\label{tw3}
&&K_{32}=1-\langle\hat{M_{2}}\rangle +  \langle\hat{M_{2}}\hat{M_{3}}\rangle - \langle\hat{M_{3}}\rangle  \geq 0\\
\label{tw4}
&&K_{34}=1+\langle\hat{M_{2}}\rangle - \langle\hat{M_{2}}\hat{M_{3}}\rangle - \langle\hat{M_{3}}\rangle  \geq 0
\end{eqnarray}
\end{subequations}
To avoid confusion,  we note here that we have changed  $\leq$ sign in Eq. (\ref{eq1}) to $\geq$ (a cosmetic change) in Eqs. (\ref{tw1})- (\ref{tw4})  and the purpose of it will be made clear shortly. As discussed in \cite{halli16,halli17}, the inequalities  (\ref{tw1}-\ref{tw4}) provides necessary and sufficient conditions for macrorealism in two-time LG scenario that is being considered here. Also, in three-time LG scenario, the inequalities (\ref{tw1})- (\ref{tw4}) along with eight more two-time LGIs provides necessary and sufficient conditions for a weaker form of macrorealism \cite{halli17}. 
 
Now, the quantum values of the LG expressions in two-time scenario given by Eqs. (\ref{tw1})- (\ref{tw4}) are derived for the state $|+m_1\rangle$. This can be linked with the two weak values. We show that there is one-to-one correspondence with the weak value and the violation of two-time LGIs. Such a connection was first pointed out in \cite{wvlgi}. But, there is an important conceptual difference between \cite{wvlgi} and our scheme.  This is due to the fact that in our scheme the system itself serves as the apparatus and no weak coupling is needed to obtain the weak value. It naturally appears in a path-only interference experiment exhibiting destructive interference. 

By writing $\hat{M_{3}}=2|+m_3\rangle\langle +m_3|-\mathbb{I}=\mathbb{I}-2|-m_3\rangle\langle -m_3|$, one can cast left hand sides of Eqs.(\ref{tw1})-(\ref{tw4}) as

\begin{subequations}
\begin{eqnarray}
\label{tww1}
&&(K_{31})_{Q}= 2 \ p(+m_3) \left[1- (\hat{M_{2}})_{w}^{|+m_3\rangle}\right] \\
\label{tww2}
&&(K_{32})_{Q}= 2  \ p(+m_3)\left[1+ (\hat{M_{2}})_{w}^{|+m_3\rangle}\right]\\
\label{tww3}
&&(K_{32})_{Q}= 2  \ p(-m_3)\left[ 1-(\hat{M_{2}})_{w}^{|- m_3\rangle}\right]\\
\label{tww4}
&&(K_{34})_{Q}= 2  \ p(-m_3)\left[1+ (\hat{M_{2}})_{w}^{|-m_3\rangle}\right]
\end{eqnarray}
\end{subequations}
where 
\begin{align}
	(\hat{M_{2}})_{w}^{|\pm m_3\rangle}= \frac{\langle +m_1|\hat{M_{2}}|\pm m_3\rangle}{\langle +m_1|\pm m_3\rangle}
\end{align}
 is the weak value of $\hat{M_{2}}$ given the pre-selected and post-selected states $|+m_1\rangle$ and $|\pm m_3\rangle$ respectively, and $p(\pm m_3)=|\langle +m_1|\pm m_3\rangle|^{2}$ is the post-selection probability.

It can be seen that the inequalities (\ref{tw1}) - (\ref{tw4}) can only be violated if the corresponding weak values  are anomalous i.e., beyond $\pm 1$. The violations of (\ref{tw1}) - (\ref{tw2}) require   $(\hat{M_{2}})_{w}^{|+m_3\rangle}>1$ and $(\hat{M_{2}})_{w}^{|+m_3\rangle}<-1$ respectively. Similarly, the inequalities of (\ref{tw3}) - (\ref{tw4}) are violated if $(\hat{M_{2}})_{w}^{|-m_3\rangle}>1$ and $(\hat{M_{2}})_{w}^{|-m_3\rangle}<-1$ respectively.  However, both $(\hat{M_{2}})_{w}^{|+m_3\rangle}$ and $(\hat{M_{2}})_{w}^{|-m_3\rangle}$ cannot be anomalous simultaneously and hence only one of the inequalities in (\ref{tw1}) - (\ref{tw4}) can be violated. In the following we demonstrate that in any path inference experiment exhibiting destructive interference implies that there exists an associated anomalous weak value.  This, in turn, provides the violation of one of the two-time LGIs given by (\ref{tw1})- (\ref{tw4}) in an interference experiment.

\section{Interference experiment, anomalous weak value and LGI} 
While our argument is valid for any path interferometric experiment, in this work, we consider the archetypical example of Mach-Zehender (MZ) setup (Fig.1). MZ set-up consists of two $50:50$ beam splitters ($BS_{1}$ and $BS_{2}$), two mirrors ($M$ and $M^{\prime}$), a phase-shifter $(PS)$ and two detectors ($D_{1}$ and $D_{2}$). The system having state $|\psi_{i}\rangle$ incidents on $BS_{1}$ where $|\psi_{i}\rangle=\alpha |\psi_{1}\rangle + \beta|\psi_2\rangle$. For simplicity, we take $\alpha$ and $\beta$ are real, satisfying $\alpha^2 +  \beta^2=1$. 

\begin{figure}[ht]
\includegraphics[width=0.5\textwidth]{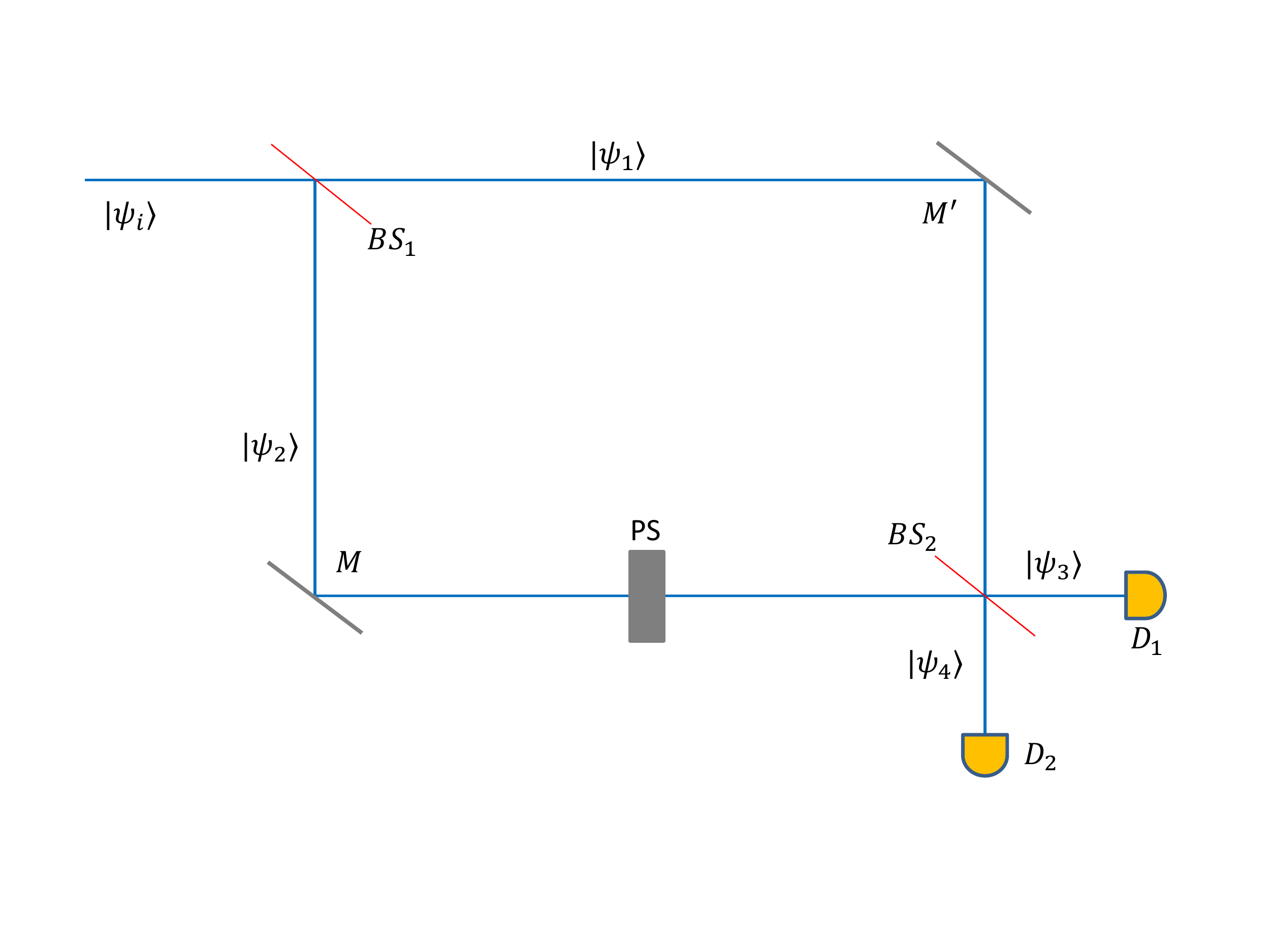}
\vskip -0.9cm
\caption{The Mach-Zehender interferometer (See text for details).}
\end{figure}
After the $BS_{1}$, the state becomes $|\psi_{BS_{1}}\rangle=\alpha |\psi_{1}\rangle +i \beta|\psi_2\rangle$ which incidents on phase-shifter (PS) and then $BS_{2}$. The state after $BS_{2}$ can be written as $|\psi_{BS_2}\rangle=\big[(\alpha+\beta) |\psi_{3}\rangle +i (\alpha-\beta) |\psi_4\rangle\big]/\sqrt{2}$ where $|\psi_3\rangle=(|\psi_1\rangle + |\psi_2\rangle)/\sqrt{2}$ and $|\psi_4\rangle=(|\psi_1\rangle -|\psi_2\rangle)/\sqrt{2}$. 

The probability of detecting the particles in the path $|\psi_{3}\rangle$ (detected at $D_{1}$) and in the path $|\psi_{4}\rangle$ (detected at $D_{2}$) are respectively given by

\begin{align}
\label{prob1}
	p(\psi_3)=\frac{(\alpha+\beta)^2}{2}; \ \ 	p(\psi_4)=\frac{(\alpha -\beta)^2}{2}
\end{align}
 We shall shortly show that when there is a destructive interference (small detection probability in one of the detectors) there exists an anomalous weak value. For example, small detection probability in the path $|\psi_{4}\rangle$ (or $|\psi_{3}\rangle$ ) indicates the existence of very large weak value. 

There is close resemblance between the weak measurement procedure and standard path interference experiment. Both of them requires a three-step procedure as follows. A pre-selection procedure to prepare the state of the system $|\psi_{i}\rangle$, interaction of the system with the interferometric set-up (state passes through $BS_{1}$) and the post-selection of particles in suitable states  ($|\psi_{4}\rangle$ and $|\psi_{3}\rangle$ by $BS_{2}$). Here, the first beam-splitter $(BS_1)$ corresponds to the measurement of dichotomic path observable $M_2=|\psi_1\rangle\langle\psi_1|-|\psi_2\rangle\langle\psi_2|$ and second beam-splitter $(BS_2)$ along with phase-shifter (PS) implements the measurement of $M_3=|\psi_4\rangle\langle\psi_4|-|\psi_3\rangle\langle\psi_3|$. 

Now, the weak values of path observable $M_2$ corresponding to the post-selected states $|\psi_3\rangle$ and $|\psi_4\rangle$ are respectively given by  
\begin{align}
	(M_{2})^{|\psi_3\rangle}_{w}=\frac{(\alpha-\beta)}{(\alpha+\beta)}; \ 	(M_{2})^{|\psi_4\rangle}_{w}=\frac{(\alpha+\beta)}{(\alpha-\beta)}
\end{align}
 with respective post-selected probabilities given by Eq. (\ref{prob1}).

It is straightforward to check that whenever $\alpha\neq\beta$ either $(M_{2})^{|\psi_3\rangle}_{w}$ or $(M_{2})^{|\psi_4\rangle}_{w}$ becomes anomalous. Thus the appearance of  the destructive interference implies the existence of anomalous weak value. However,  $(M_{2})^{\psi_3}_w$ and $(M_{2})^{\psi_4}_{w}$ both cannot be anomalous together. In extreme condition, when $\beta=0(1)$ and $\alpha=1(0)$, there is no destructive interference and both the weak values  $(M_{2})^{|\psi_3\rangle}_{w}=(M_{2})^{|\psi_4\rangle}_{w}=1$ with same post-selection probability $p(\psi_{3})= p(\psi_{4})=1/2$. There is an important difference between the path weak value considered here and the standard weak value \cite{aav}. In our scheme, no additional apparatus is involved and the system itself acts as an apparatus. If the pre-selected state of the system is taken to be a pure state, it  remains in a pure state after the interaction with the $BS_1$. So, no explicit weak coupling for the measurement at $BS_1$ is needed to be ensured. 

We note here that there is an intense debate \cite{ferrie,qin} whether weak values are inherently quantum or rather a purely statistical feature of pre- and post-selection with disturbance. The view that classical probabilities with suitable noise mimic weak value has been criticized by Qin \emph{et al.} \cite{qin} through the Stern-Gerlach setup for spin measurements. It is demonstrated by Pusey \cite{pusey} that anomalous weak values are proof of contextuality which is experimentally verified in \cite{pia}. Dressel \cite{dressel} argued that weak values arises due to quantum interference effect. 

Here, we argue that the converse is also hold, i.e., for any quantum interference experiment there is an associated anomalous weak value of path observable. However, there is an important difference. In \cite{qin,dressel}, the interference of the post-selected apparatus (not the system) states plays a key role in contrast to our scheme where no separate apparatus is involved. 
 \vskip 0.2cm
\begin{figure}[ht]
\includegraphics[width=0.5\textwidth]{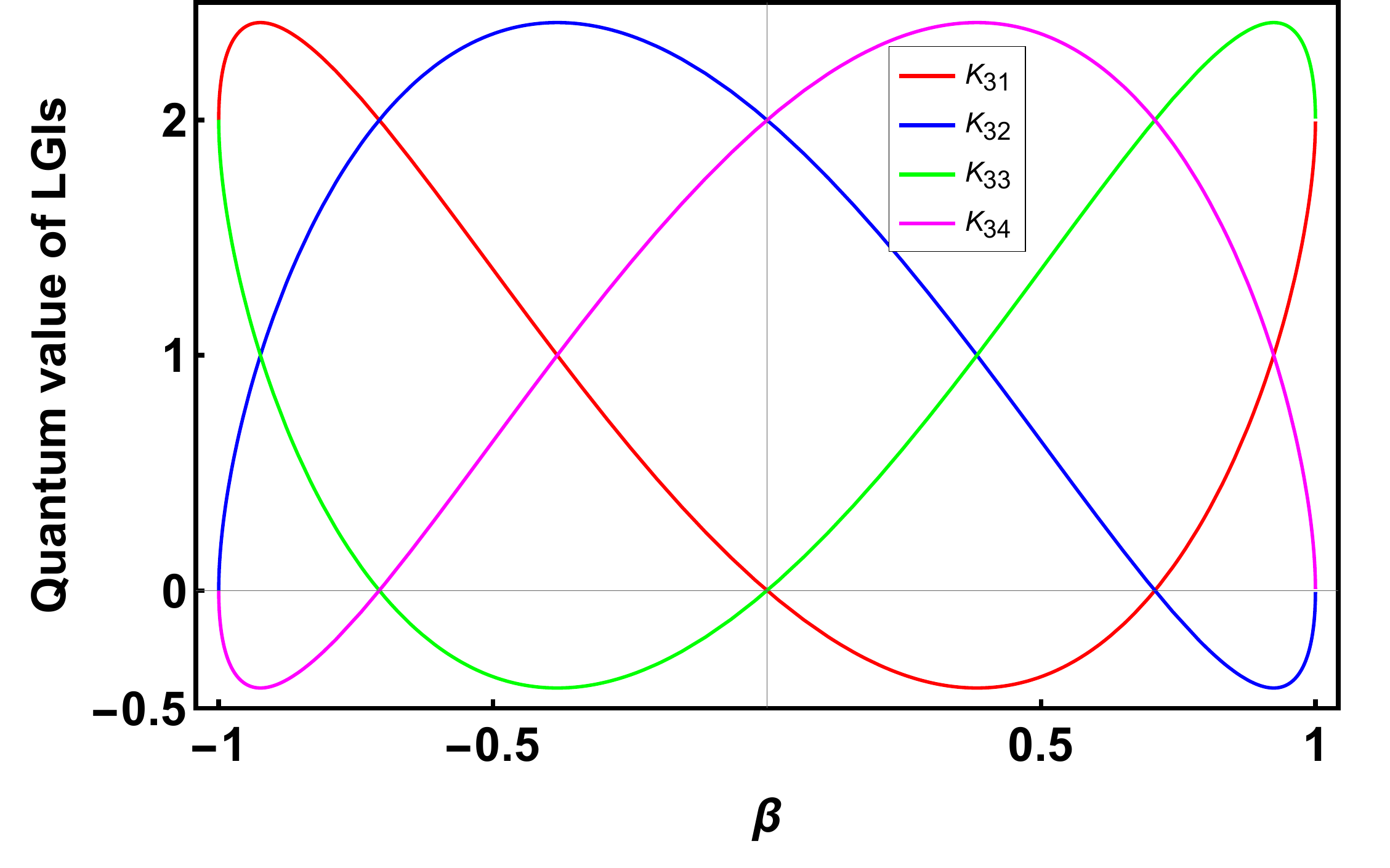}
\vskip -0.3cm
\caption{Quantum values of four LG expressions given in Eqs. (\ref{k31}) - (\ref{k33}) are plotted against $\beta$. Curves demonstrate that one of the four LGIs will be violated for any given value of $\beta$, except for $\beta= 0$ or $\pm 1$.}
\end{figure}

The arguments provided up to now  thus enables us to demonstrate that if the path interference experiments exhibits destructive interference then there is a violation of LGIs. In other words, the interference experiment itself constitute a proof of incompatibility between QM and realism through the violation of LGIs. In order to explicitly show this, by identifying $|-m_3\rangle=|\psi_{3}\rangle$, $|+m_3\rangle=|\psi_{4}\rangle$ and $|+m_1\rangle=|\psi_{i}\rangle$, we can write Eqs.(\ref{tww1})-(\ref{tww4}) as

\begin{subequations}
\begin{eqnarray}
\label{k31}
(K_{31})_{Q}= 2\beta (\beta -\alpha); \ \ \ (K_{32})_{Q}= 2\alpha (\alpha-\beta)\\
\label{k33}
(K_{33})_{Q}= 2\beta (\alpha +\beta); \ \ \ (K_{33})_{Q}= 2\alpha (\alpha +\beta)
\end{eqnarray}
\end{subequations}
For the values, $\alpha>\beta$, from Eq. (\ref{tww1}), we have $(K_{31})_{Q}<0$ implying the violation of LGI in (\ref{tw1}) and consequently $(M_{2})^{|\psi_4\rangle}_{w}>1$ (i.e., anomalous weak value).  If $\beta>\alpha$, we have the violation of LGI in  (\ref{tw2}) and $(M_{2})^{|\psi_4\rangle}_{w}<-1$.  In both the above cases, we have destructive interference at the detector $D_2$. Similar arguments can be made for Eq. (\ref{k33}) which requires anomalous values of $(M_{2})^{|\psi_3\rangle}_{w}$.  Hence, the existence of anomalous weak value in a path-only interference experiment warrants the violation of two-time LGIs. One may then argue that the interference experiments for large molecule \cite{arndt,gerlich} have already tested the quantum violation of macrorealism proposed by LG.    

\section{Non-invasive measurability, two-time LGIs and macrorealism}
Since the introduction of LGIs it remains a debatable issue  regarding what specific notion of macrorealism is tested in LG scenario. The roots the assumptions made in LGIs is explained by Leggett in many of his writings, for example, see \cite{leggett02}, This is significantly motivated from Bohr's  `solution'  of quantum measurement problem ( denying definite properties to microscopic objects in the absence of observation but asserts  macroscopic objects to have such properties at all times irrespective of observation) and Schr$\ddot{o}$edinger's  criticism to such insistence (there should not be any logical reason to insist that an electron and an everyday object behaves differently in quantum theory unless a well-defined criteria is set by theory itself). He argued that it is not always necessary to have quantum description of everyday objects but, in principle, it should be legitimate to ask for a quantum mechanical account for them. 

In LG scenario, there are two subtle issues regarding the assumptions used to derive LGIs. For an excellent review with enticing discussion, we refer \cite{maroney}.  First,  what exactly the `state' implies in MRps assumption  in a macrorealistic theory is unclear. A macrorealist would have desired for ontic states. The macrorealistic model advocated by Leggett and Garg strongly suggests that its background framework is in fact quantum theory along with the superselection rule for denying the linear superposition in our eveyday world. If macrorealism is understood as the realism of the macroscopic system, then an one-to-one correspondence between macroscopically distinct states and definite values of macro object  is needed to be ensured, irrespective of observation. It is the fundamental feature of quantum theory that the states do not directly corresponds to the properties unless it is a classical mixture of eigenstates of the observables. 

Note that, the ontological model of quantum theory put forwarded by Bohm \cite{bohm}, being statistically equivalent to quantum theory, ensures the deterministic properties of  the system of undivided world by keeping the linear superposition. However, there is a subtle difference between deterministic and definite value within Bohmian model. One may argue that determinism is a weaker notion than definite value as later implies the former but converse is not true. As argued by Vaidman \cite{vaidman} that Bohmian positions (the ontic states) of a given ensemble are fixed by the modulus of wave function and the outcomes of measurements of the observables are predetermined, but may not be value definite. This is due to the fact that Bohmian model is inherently contextual as different experimental setups for measurements of the same observable may lead to different observed values. Such a contextuality is not similar to the Kochen-Specker \cite{kochen} or Spekkens \cite{spek05} form.  Then the Bohmian model is deterministic, but contextual value definite. It may then be assumed that the `state' in MRps assumption involved in LG scenario refers to the ontic state providing non-contextual definite value.  

A few more comments would be helpful before concluding this part of discussion. An elegant and generalized framework of ontological model was proposed by Spekkens \cite{spek05,hari} that corresponds to any operational theory and without reference to quantum theory.  For example, the notion of non-contextuality  introduced in \cite{spek05} is not based on quantum theory in contrast to the Kochen-Specker version.  But, original LG formulation is so tied up with quantum theory that it is difficult to provide a model independent formulation by separating it out from the conceptual influence of quantum framework. 

Now we come to the second issue, i.e., the non-invasiveness measurability (NIM) assumption in a macrorealistic model. This issue remains debatable and most discussed in the literature. Leggett's original view regarding NIM is based on the ideal negative-result measurement in QM and he argued that it is a natural corollary of the MRps. Note that, the statistical version of NIM condition - the operational non-invasivness (also well-known as no-signaling in time condition ), implies that prior measurements do not influence the statistics of future measurements.  In two-time and three-time LG scenario they imply LGIs \cite{clemente15,clemente16,halli17,maroney}. Mathematically,

\begin{align}
	p(m_j)=\sum\limits_{m_i} p(m_i, m_j)
\end{align}
meaning that the prior measurement does not disturb the subsequent measurement when $i<j$. In \cite{clemente15,clemente16}, Clemente and Kofler argued that a suitable set of such operational non-invasiveness conditions provide the necessary and sufficient condition (NSC) for macrorealism for three-time LG scenario, while three-time LGIs do not. Interestingly, in two-time LG scenario,  the four inequalities given by (\ref{tw1})-(\ref{tw4}) provide the NSC for macrorealism. 

In general, the operational non-invasivness condition is not  satisfied in QM. However, as argued in \cite {maroney} that the NIM is an independent condition than MRps in contrast to Leggett's view. Thus what the violation of LG signifies is unclear. A macrorealist may claim that it is simply due to the violation of operational non-invasiveness and nothing can be said about the violation of macrorealism \emph{per se}. Note that, NIM condition can also be assumed as ontic non-invasiveness as in Spekkens \cite{spek05} formulation. This is a stronger reading of NIM condition than operational non-invasiveness. While the former implies later but converse does not hold \cite{maroney}. Even in such case, to provide the NIM condition the  similar status of locality in Bell scenario, the operational non-invasiveness in QM still needs to be satisfied. As mentioned earlier, there has been a few proposals to achieve this goal. One particular approach is by using quasiprobability proposed by Halliwell \cite{halli16}. Using this approach we demonstrate how operational non-invasiveness is automatically satisfied in the interference experiment but LGIs are violated.   

Note that LG test requires the joint sequential probabilities of two non-commuting observables and there is no unique prescription is  available in QM. In \cite{halli16}, it is proposed that instead of $p(m_i, m_j)$ one can use a suitably defined quasiprobabilities $q(m_i, m_j)$. Importantly, they correctly reproduce the sequential correlation and crucially, satisfy the operational non-invasivness in QM. Such quasiprobabilities are defined as 

\begin{align}
	q_{Q}(m_i ,m_j) =\frac{1}{2}Tr[\left\{P_{m_j}P_{m_i}	+P_{m_i}P_{m_j} \right\}\rho]
\end{align}
satisfying $\sum\limits_{m_i, m_j} q_{Q}(m_i ,m_j)=1$ where $P(m_{i(j)})= (\mathbb{I}+m_{i(j)}M_{i(j)})/2$. Due to the symmetry in order, they satisfy operational non-invasiveness in QM,

\begin{subequations}
\begin{eqnarray}
\label{oi1}
	p(m_j)=\sum\limits_{m_i}	q(m_i, m_j)= Tr[P(m_j)\rho] \\
	\label{oi2}
	p(m_i)=\sum\limits_{m_j}	q(m_i, m_j) = Tr[P(m_i)\rho]
\end{eqnarray}
\end{subequations}
This means that $p(m_j)$ remains independent of the fact if a prior measurement of $M_i$ is performed. Importantly, the correlation remains same as  
\begin{align}
	\langle M_{i} M_{j}\rangle=\sum\limits_{m_i, m_j}m_i m_j \ q(m_i ,m_j)=\sum\limits_{m_i, m_j}m_i m_j  \ p(m_i, m_j)
\end{align}

The macrorealistic reading of the quasiprobabilities is given by  
\begin{eqnarray}
	q(m_i, m_j)= \frac{1}{4}(1+ m_i \langle M_{i}\rangle +m_j \langle M_{j}\rangle+m_i m_j\langle M_{i} M_{j}\rangle
\end{eqnarray}
The quasiprobabilities $q(m_i ,m_j)$ can be negative or positive. Importantly,  $q(m_i ,m_j)>0$ implies the four two-time LGIs (with a multiplicative factor 4) and also provide the NSC for macrorealism in two-time LG scenario. As argued in \cite{halli16,halli17}, in three-time LG scenario when $i,j=1,2,3$ with $j>i$, twelve such quasiprobabilities provide NSC for a weaker form of macrorealism.  Thus, in our case $q(m_i , m_j)<0$ implies the violation of two-time LGIs. But, since operational non-disturbance remains satisfied, such a violation constitutes a violation of both the MRps and ontic non-invasiveness.    

 In our interferometric experiment, if we identify $|+m_{2}\rangle\equiv |\psi_{1}\rangle$, $|-m_{2}\rangle\equiv |\psi_{2}\rangle$ $|+m_{3}\rangle\equiv |\psi_{3}\rangle$ and $|-m_{3}\rangle\equiv |\psi_{4}\rangle$, then  by using Eqs. (\ref{oi1}) -(\ref{oi2}) the probabilities $p(\psi_3)$ and $p(\psi_4)$ can be written  as 

\begin{subequations}
\begin{eqnarray}
	p(\psi_3)=q(\psi_1, \psi_3)+q(\psi_2, \psi_3)\\
	p(\psi_4)=q(\psi_1, \psi_4)+q(\psi_2, \psi_4)
\end{eqnarray}
\end{subequations}
where $q(\psi_1, \psi_3)$ is one of the four quasiprobabilities. It is simple to check that in QM, the value of $p(\psi_3)$ and $p(\psi_4)$ match with Eq. (\ref{prob1}), and thus satisfying the operational non-invasiveness. As mentioned, the four conditions $q(\psi_1, \psi_3)>0$ (with a multiplicative factor 4) are just the four two-time LGIs in \ref{tw1}-\ref{tw4}. As already shown (see also, Fig.2) that one of the LGIs given in \ref{tw1}-\ref{tw4} will be  violated whenever destructive interference occurs. This explains how the  no-signalling in time condition is satisfied in QM but LGI is violated. Our scheme thus provides a conclusive test of macrorealism in QM.   
     
\section{Discussion} 
In sum, we have provided an hitherto unexplored link between the interference experiment and incompatibility of macrorealism in QM through the violation of LGI. The interference experiment for large objects \cite{arndt,gerlich} is a practical approach to test the macroscopic quantum coherence.  On the other hand, LG formulation  is a conceptual approach for testing the incompatibility of the notion of macrorealism in QM in our everyday world. In this work, we demonstrated that whenever there is destructive interference there exists an anomalous weak value that enables one to demonstrate the quantum violation of LGIs. Further, we have provided a detail discussion regarding the assumptions involved in LG framework and how our scheme fits into that framework.

The LGIs are derived based on two key assumptions; macrorealism \emph{per se} and non-invasive measurability. Since LGI involves the measurements of non-commuting observables, then a prior measurement in general disturbs the future measurements and hence statistical version of non-invasive measurability (the no-signalling in time) is not in general satisfied in QM. In such a case, the quantum violation of LGIs may not convince a macrorealist  who may claim to salvage the macrorealism \emph{per se} by simply abandoning the non-invasive measurability. Hence, unless this loophole is closed in experiment the LG test of macrorealism is \emph{not} as conclusive as the test of local realism through Bell's inequalities. 

There have been a few interesting proposals in the literature regarding how to close this loophole. In their original work, Leggett-and Garg \cite{leggett02}advocated the negative-result measurement of QM to validate the non-invasive measurability which was later tested in experiment \cite{goggin}. Proposals using weak measurement \cite{avella}, quasiprobabilities \cite{halli16} and continuous velocity measurements \cite{hali16aa, maji} have also been reported. But, a string of criticisms \cite{maroney} have also been made regarding the viability of some of the approaches. It is argued that in negative result measurement the collapse from distance occurs and subsequent dynamics of the state is disturbed. In LGI test through the weak measurement of spin or polarization observable, it is assumed that the system and apparatus is minimally entangled but in principle invasiveness still occurs due to this entanglement.  

In contrast, our scheme does not involve measurement apparatus (the root of causing disturbance by entangling the system) and system itself serves the role of apparatus. Thus, the non-invasive measureability assumption is in principle satisfied. In order to explicitly explain it, we use an interesting approach involving quasiprobabilities. Such probabilities mimics the sequential LG correlation and crucially satisfies the no-signalling in time condition in our two-time LG scenario. Moreover, the quasiprobabilities, while positive, are proportional to four two-time LGIs providing the necessary and sufficient condition for macrorealism in two-time LG scenario. It is shown that unless the quantum superposition exists, one of the quasiprobabilities are negative implying the violation of LGIs and consequently the weak value becomes anomalous. We can then infer that the scheme presented in this paper, one cannot salvage the macrorealism in QM by simply abandoning the non-invasive measurability. This thus provides a conclusive test of the LG formulation of macrorealism.

\acknowledgments
 AKP acknowledges the support from Ramanujan Fellowship research grant (SB/S2/RJN-083/2014). 

\end{document}